# Adaptive Prognostic Malfunction Based Processor for Autonomous Landing Guidance Assistance System Using FPGA

**Hossam O. Ahmed[1,3] (Senior, IEEE), and David Wyatt[2]**
[1]College of Engineering and Technology, American University of the Middle East, Kuwait
[2]Pixeldisplay, USA
[3]Corresponding author

**ABSTRACT** The demand for more developed and agile urban taxi drones is increasing rapidly nowadays to sustain crowded cities and their traffic issues. The critical factor for spreading such technology could be related to the safety criteria that must be considered. One of the most critical safety aspects for such vertical and/or Short Take-Off and Landing (V/STOL) drones is related to safety during the landing stage, in which most of the recent flight accidents have occurred. This paper focused on solving this issue by proposing decentralized processing cores that could improve the landing failure rate by depending on a Fuzzy Logic System (FLS) and additional Digital Signal Processing (DSP) elements. Also, the proposed system will enhance the safety factor during the landing stages by adding a self-awareness feature in case a certain sensor malfunction occurs using the proposed Adaptive Prognostic Malfunction Unit (APMU). This proposed coarse-grained Autonomous Landing Guidance Assistance System (ALGAS4) processing architecture has been optimized using different optimization techniques. The ALGAS4 architecture has been designed completely using VHDL, and the targeted FPGA was the INTEL Cyclone V 5CGXFC9D6F27C7 chip. According to the synthesis findings of the INTEL Quartus Prime software, the maximum working frequency of the ALGAS4 system is 278.24 MHz. In addition, the proposed ALGAS4 system could maintain a maximum computing performance of approximately 74.85 GOPS while using just 166.56 mW for dynamic and I/O power dissipation.

## I. INTRODUCTION

As the world population continues to grow, it will become almost impossible to rely on existing transportation systems to cope with the growing demand. Air pathways allocated at different levels have been the subject of many attempts and proposals. Therefore, it is imperative to intensify the precautions and standards for autonomous landing safety for the next generation of Urban Air Mobility (UAM) from several different angles [1-3]. It should also be taken into consideration that the new generation of UAM drones will depend mainly on their vertical takeoff and landing (V/STOL) capability. The embedded V/STOL mechanisms must be highly effective against unexpected and likely challenges, such as inept pilots and limited and/or unexpected landing sites during emergencies due to the enormous UAM taxi drones that are expected to cover each city in the near future [4-5]. Improving the efficiency of autonomous V/STOL mechanisms will also benefit related space and military applications [6]. When the following subsidiaries are included, this issue becomes a formidable obstacle. First, it is crucial to be aware that about 53.85% of all contemporary flight accidents occur during the final three phases of a flight, which comprise the Initial Approach, Final Approach, and Landing phases of an airplane [7, 8]. Second, it is necessary to presume that the envisaged UAM technology will be based on either semi or fully autonomous flying systems. These UAM drones are anticipated to be very sophisticated and intricate. Thus, it will be almost impossible to afford the predicted demands for such a large number of skilled pilots to cover a certain small geographical area. Third, the intended UAM technology must account for the fact that the routing of these route(s) per trip will be dynamically allocated and adaptable depending on a range of elements and situations that may spontaneously arise [4, 9, 10]. These circumstances could be natural or manmade and might trigger the activation of emergency and precautionary control measures. Only with these restricted factors might we have a clearer picture of how these taxi drones need to be made safer.

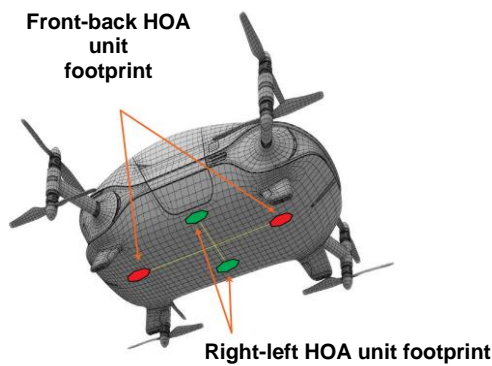

FIGURE 1. **The graphical illustration of the proposed constellation of the two pairs of the differential HOA units.**

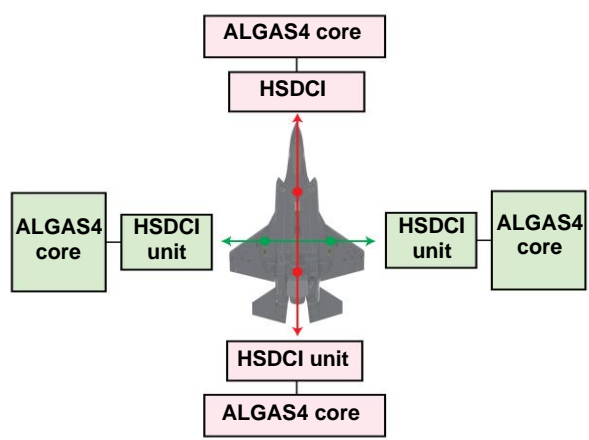

FIGURE 2. **The proposed distributed processing units of a complete ALGAS4 system [5].**

Consideration should also be given to the fact that the majority of these taxi drones will rely on the Vertical and/or Short Take-Off and Landing (V/STOL) mechanism, which will increase the complexity of the electronic components required to ensure the desired safety parameters [11]. Several researchers contribute to the improvement of all flight stages' safety measures by adopting modern high-tech approaches such as Machine Learning, Deep Neural Networks (DNN), and other sophisticated control systems such as the Fuzzy Logic System (FLS) [12].

Also, researchers are working to find novel simulation tools to enable safe and efficient autonomous on-demand free flight operations within a metropolitan area [13]. This objective might be realized by designing decentralized and collaborative computing cores to operate the autonomous landing mechanism by collecting sensory input from the distributed Hybrid Obstacle Avoidance (HOA) sensory nodes mounted on the underside of a drone, as illustrated in Fig. 1. The proposed processing architecture for the Autonomous Landing Guidance Assistance System (ALGAS4) could provide a real-time solution for preventing landing hazards in various types of aircraft and autonomous drones.

The proposed ALGAS4 system is composed of four ALGAS4 processing cores that are decentralized. Each ALGAS4 processing core is primarily dependent on two processing modules: coarse-grained Fuzzy Logic System (FLS) processing cores and the Adaptive Prognostic Malfunction Unit (APMU). In addition, it relies on additional Digital Signal Processing (DSP) substances, as will be outlined in the sections that follow. In Section II, we explore related works, comparable research, and similar contributions. In Section III, we discuss the potential of the proposed ALGAS4 design. In Section IV, we presented the experimental findings derived from the ALGAS4 architectural synthesis process utilizing Intel Quartus Prime. In Section V, we discussed the conclusion and future work.

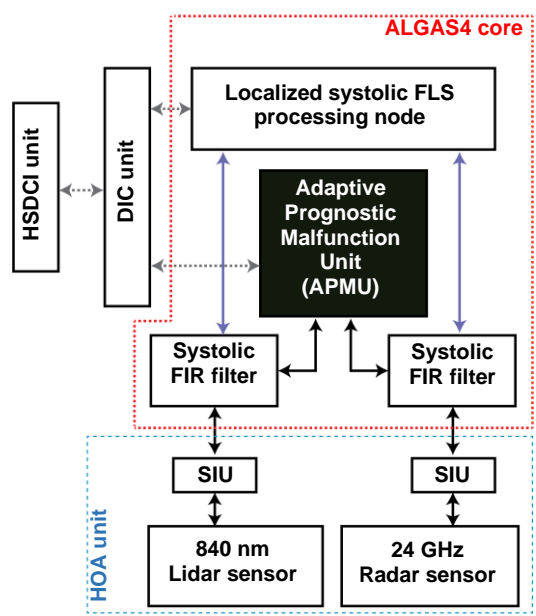

FIGURE 3. **The detailed structure of one spatial corner of the proposed ALGAS4 system.**

## II. Related work

To ensure the success of the application of UAM worldwide, it is paramount to have a well-developed understanding and a considerable degree of efficiency when applying safety factors. New and adaptable safety standards should be integrated with the existing safety regulations applicable to the Single European Sky ATM Research (SESAR) and Next Generation (NextGen) air transportation systems [14, 15]. Different research studies explained the importance of applying advanced control systems and Artificial Intelligence (AI) algorithms to enhance the variety of capabilities of UAM systems [16-18]. Several proposals have been developed related to trajectory conflict detection and avoidance mechanisms, smart path planning, and configurable air traffic control approaches [19-21].

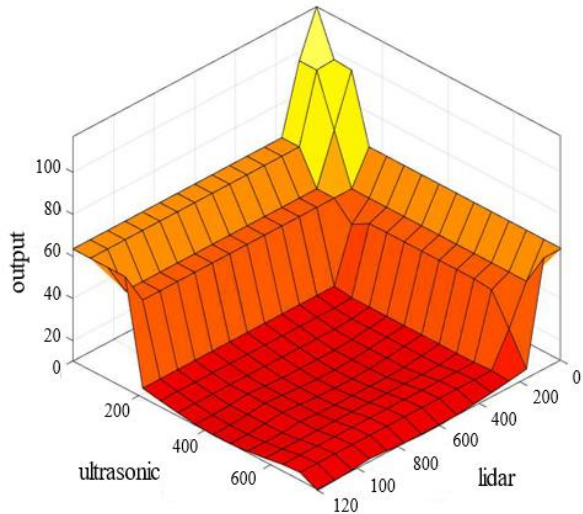

FIGURE 4. **The surface graph of the proposed Open FLS processing core.**

Many research contributions are using image processing and integrated computer vision to strengthen the landing safety of drones. Also, Reference [22] outlines a method and design for a system that manages multiple data streams through a multicore asymmetric processing architecture, aiming to eliminate data interruptions directed at the application processors. This design specifically supports a controlled environment for flight software within NASA's Safe and Precise Landing – Integrated Capabilities Evolution (SPLICE) project. SPLICE focuses on advancing sensor, algorithm, and computing technologies for Precision Landing and Hazard Avoidance (PL&HA) capabilities. The computing technology used in this paper, known as the Descent and Landing Computer (DLC), houses several SPLICE algorithms demanding high computational resources, requiring real-time and deterministic execution. Operating on a custom Single Board Computer (SBC) featuring a Xilinx Ultrascale+ Multiprocessor System-on-a-Chip (MPSoC), the software receives input data from diverse sensors with varying data rates and packet sizes.

A data pathway is devised between SPLICE sensors and algorithms, efficiently delivering this data to the flight software using the MPSoC's asymmetric processing cores and Field Programmable Gate Array (FPGA) fabric. This setup isolates the application processors executing the flight software from the interruptions associated with incoming data. By capitalizing on real-time processors on the MPSoC and a structured interface within the shared memory on the SBC, the flight software can optimally utilize the complete set of application processors. Each processor's available capacity within this set is maximized for SPLICE applications, ensuring a sufficiently deterministic execution environment without the complexities and overheads of a real-time operating system. Reference [23] suggests an Unmanned Aerial Vehicle (UAV) positioning system utilizing a landing platform equipped with four ultra-wideband (UWB) anchors positioned on the platform itself, along with two tags and an Inertial Measurement Unit (IMU) installed on the UAV. This system offers enhanced accuracy in UAV localization during landing. In contrast to traditional visual localization systems, this landing platform-based UAV localization system remains unaffected by lighting conditions like smoke, eliminating reliance on the Global Positioning System (GPS). Moreover, the compact space needed for anchor deployment allows convenient installation on unmanned vehicles. The trial outcomes demonstrated that this localization system achieves decimeter-level accuracy for UAV landings on the designated platform with a remarkable 61% reduction in the average RMSE (Root Mean Square Error). Nonetheless, this system doesn't maximize the utilization of data from the two UWB tags to furnish directional cues for the UAV, instead relying on altitude data provided by the UAV itself.

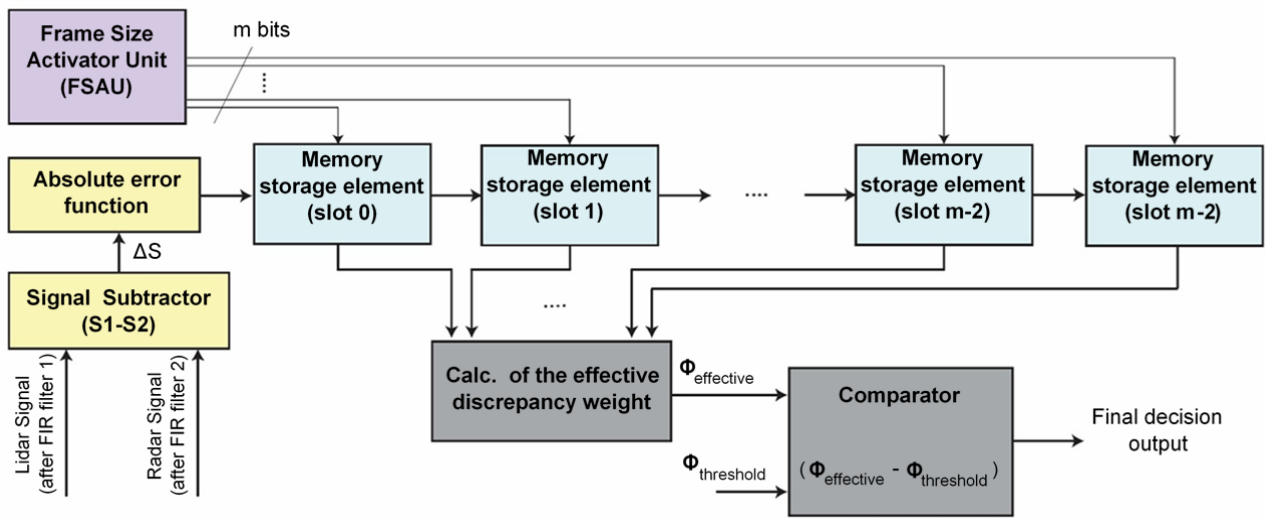

FIGURE 5. **The detailed structure of the Adaptive Prognostic Malfunction Unit (APMU).**

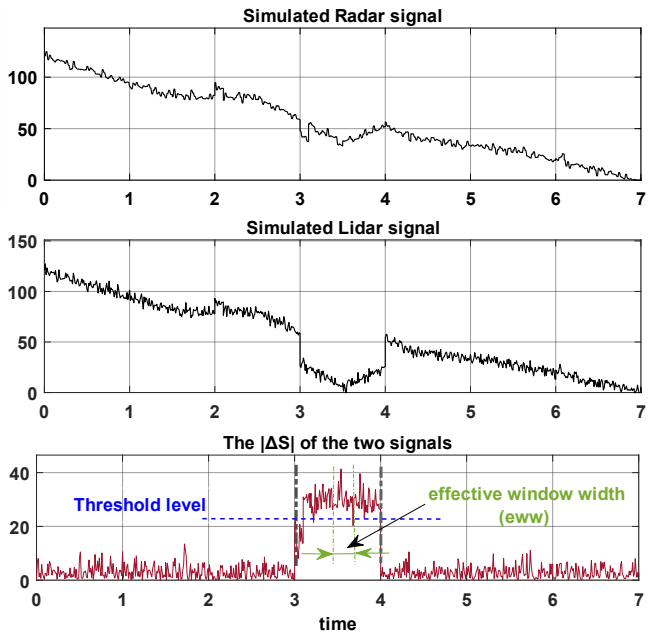

FIGURE 6. A simulated scenario of the behavior of the APMU under a certain degree of signal discrepancy from the Radar and Lidar sensors.

Another interesting study aimed to enhance the autonomous landing of a quadrotor on inclined surfaces of up to 40 degrees, solely using radar sensors and active asymmetric skids, even in high-particle environments [24]. Numerical simulations factored in perceptual errors to establish an acceptable margin, preventing contact between the propeller and the sloped surface. During quadrotor testing, successful landings were accomplished on surfaces inclined up to 40 degrees with an accuracy of within 3 degrees. The radar sensors exhibited performance comparable to, if not better than, the camera depth, even in a testing environment obscured by fog. Through this radar-based approach, the quadrotor demonstrates enhanced capabilities to handle high-particle environments and execute landings on sloped surfaces.

Nevertheless, the efficiency of autonomous landing will be one of the most critical safety concerns for the next generation of UAM systems. The major concern about landing safety arises from the anticipated wide gap and shortage in the market between the predicted abundance of UAM vehicles and the supply of qualified pilots. Also, there is the additional challenge of landing on unpredictable terrain, especially under emergency landing conditions [11, 25]. Recently, the fuzzy logic system (FLS) has been proposed as an advanced control algorithm, particularly to resolve many UAM issues, such as autopilot dynamics and visual human tracking in drone systems [26-27]. Moreover, the FLS is a remarkably robust advanced control system that can handle signal ambiguity and noise in electronic systems [28-29]. FLS has been demonstrated to optimize safety hazards in self-landing and hovering missions for UAM vehicles [30-32].

A recent contribution in this domain introduced a decentralized processing system aimed at boosting safety measures during critical phases of V/STOL drones [5]. The system incorporates various processing and control algorithms, including an integrated Open Fuzzy Logic System (FLS), Flight Rules Unit (FRU), FIR filters, and a unique Prognostic Malfunction processing unit. Through multiple optimization techniques, the design, this proposed design achieves a peak computational performance of 70.82 Giga Operations per Second (GOPS) while maintaining low thermal power dissipation of 145.4 mW, suitable for mobile avionic systems using the INTEL FPGA chip. Also, this paper aligns with evolving UAM guidelines from FAA, SESAR, and NextGen safety measures during the autonomous landing of taxi drones.

### III. PROPOSED ALGAS4 PROCESSING UNIT

This paper's major contribution is to enhance the safety profile of V/STOL taxi drones during the landing phase. By continuously measuring the distance between the taxi drone and the landing spot from four separate HOA units and processing this aggregated data using the suggested ALGAS4 system, a fully autonomous system might be developed that guarantees a reliable and effective landing procedure. As shown in Fig. 2, the proposed ALGAS4 system consists of four ALGAS4 cores functioning together to form a decentralized and collaborative processing system. Each pair of spatially opposed ALGAS4 cores creates a differential processing pair to maximize the safety factors by continuously confirming that the measured distance on the opposite sides of the taxi drone is the same and within the predefined tolerance. The High-Speed Differential Communications Interface (HSDCI) unit facilitates communications and data sharing between these ALGAS4 cores. Also, Fig. 3 depicts the detailed structure of one spatial core of the proposed ALGAS4 system.

Each ALGAS4 processing corner is comprised of four primary subparts: the HOA unit, the ALGAS4 processing core, the HSDCI, and the Differential Inclination Control (DIC) unit. The liability of the DIC unit is to accumulate and determine the important information that must be shared with other ALGAS4 processing cores from a specific ALGAS4 processing unit. In addition, it is tasked with creating the packet format containing this crucial data before delivering it to the HSDCI unit. The HSDCI unit is then responsible for determining the proper settings of the communications link, such as the link speed and flow control. The HOA unit consists of the hardware of the sensory modules and the Sensor Interface Unit (SIU). Within a single ALGAS4 processing core, we have two 15-TAPS FIR filters, one systolic FLS unit, and the APMU. Most of these building blocks have been

VOLUME XX, 2017 x

explained in previous contributions except the HSDCI, SIU, and DIC [4, 5, 11]. Each Mamdani-based model of the proposed FLS processing core depends on eleven fuzzy rules. The overall computational procedure of the proposed landing control system is depicted in Algorithm 1 as shown below. Also, Algorithm 1 shows the eleven fuzzy rules of the FLS unit. These if-else rules have been exemplified using fuzzy qualifiers (Italics) and logical semantics (in CAPS). As: E: Extremely, N: Near, M: Middle, F: Far, L: Low, H: High.

---

**Algorithm 1** The parallel processing procedure of the FLS unit in the ALGAS4 system.

---

*Initialization:*
1:   Reset all the registers to zero
2:   Keep the Fuzzy Logic Engine deactivated till a reasonable input signal activity is recorded.

*Step 1: Aggregating the radar and lidar sensor values in parallel*
3:   **if** (the radar sensor is sending a new signal activity) **then**
4:     store the new sample and activate the fuzzifier unit.
5:   **end if**
6:   **if** (the lidar sensor is sending a new signal activity) **then**
7:     store the new sample and activate the fuzzifier unit of the FLS
8:   **end if**

*Step 2: Activate all the membership function units in parallel*
9:   **if** (the fuzzifier unit is activated) **then**
10:     all the membership function units will fire their output values in parallel to the FLS Inference Engine
11:   **end if**

*Step 3: Apply all the if-else of the FLS in parallel*
12:   **if** (lidar is EN) **and if** (Radar is EN) **then** (Output_to_Drone central processor is EH) **end if**
13:   **if** (lidar is N) **and if** (Radar is EN) **then** (Output_to_Drone central processor is H) **end if**
14:   **if** (lidar is EN) **and if** (Radar is N) **then** (Output_to_Drone central processor is H) **end if**
15:   **if** (lidar is N) **and if** (Radar is N) **then** (Output_to_Drone central processor is H) **end if**
16:   **if** (lidar is M) **and if** (Radar is M) **then** (Output_to_Drone central processor is M) **end if**
17:   **if** (lidar is F) **and if** (Radar is M) **then** (Output_to_Drone central processor is M) **end if**
18:   **if** (lidar is F) **and if** (Radar is F) **then** (Output_to_Drone central processor is L) **end if**
19:   **if** (lidar is EF) **and if** (Radar is F) **then** (Output_to_Drone central processor is L) **end if**
20:   **if** (lidar is F) **and if** (Radar is EF) **then** (Output_to_Drone central processor is L) **end if**
21:   **if** (lidar is EF) **and if** (Radar is EF) **then** (Output_to_Drone central processor is L) **end if**
22:   **if** (lidar is M) **and if** (Radar is F) **then** (Output_to_Drone central processor is M) **end if**

*Step 4: calculate the defuzzification output*
23:   **if** (the FLS Inference Engine finish signal is active) **then**
24:     activate the defuzzification unit and calculate the final crisp output value
25:     send the finish signal to the next system unit
26:   **end if**
27:   **return** to step 1

---

Also, Fig. 4 depicts the relationship between the two input sensory data and the final output of each fuzzy processing core in the proposed ALGAS4 processing core. Nevertheless, this study focused mainly on the newly proposed APMU module. The main idea of the APMU is to add more safety capabilities to the proposed ALGAS4 system. Moreover, the FLS is concerned about the control of the progressive landing procedure under the condition that the signal discrepancy between the readings of the two sensors remains within an acceptable range. Thus, it is crucial to add a new landing safety layer for the occurrence of any unsatisfactory incident, including a discernible divergence in such readings. These errors could be the result of a sensor malfunction, a cyberattack, or a jamming attack. Subsequently, the APMU unit was proposed to solve this issue. As illustrated in Fig. 5, the APMU unit calculates the Mean Absolute Error (MAE) between the two signal readings from the lidar and the radar sensors, S1 and S2, respectively, as indicated in (1). In general, the MAE is a measurement of the errors that occur between paired observations that reflect the same phenomenon.

TABLE I
THE COMPARISON BETWEEN THE PROPOSED APMU AND THE PMU IN [5]

| | Adaptive Prognostic Malfunction Unit in this paper | Prognostic Malfunction Unit in [5] |
|---|---|---|
| *Targeted FPGA device name* | INTEL 5CGXFC9D6F27C7 | |
| *Embedded FPGA's DSP resource usage* | None | |
| *Total dynamic thermal power dissipation (I/O and core)* | 13.64 mW | 11.44 mW |
| *External memory usage or BRAM* | None | |
| *Max. Frequency* | 395.73 MHz | 431.03 MHz |
| *No. of logic elements* | 153 ALMs | 46 ALMs |
| *Min. decision depth* | 4 Soft memory slots | 1 Soft memory slot (accumulated values) |
| *Max. event window size to affect the final output* | 16 Soft memory slots | 1 Soft memory slot (accumulated values) |
| *Effective window width (eww)* | Adjustable (1 to 16 samples) | Fixed window size (16 samples) |

The MAE is determined by computing the sum of all absolute errors and dividing it by the total number of samples (n). From the hardware perspective, the APMU is calculating the difference value, $\Delta S = S1-S2$, between two sensory sample data at a certain sampling time t. Then, the absolute error function unit is responsible for generating the corresponding absolute signal for each sensory sample difference $|\Delta S|$. Every $|\Delta S|$ sample is stored in a FIFO-like memory storage buffer. The size of this FIFO-like memory storage buffer is (m) cells, and it is equal to the number of signal samples (n). In this paper, the maximum value of the active memory storage element is limited to 16 units. The adaptive feature of the APMU is due to its ability to adjust what is called "the depth of decision (n)", which the drone's main processor will decide to select according to the scenario and the working environment of the flight. Hence, the Frame Size Activator Unit (FSAU) has m-bits to activate the number of memory storage elements. The minimum number of memory storage elements m is four, and the maximum number is sixteen. Also, the division operator is consuming more logic resources in the implementation process. Hence, we decided to replace the division operation by the number of data points (n) in (1), with another concept of a "subtract-compare-based" operation. Furthermore, we found that it is important to count the existence of the sample points in which a recorded discrepancy event occurs (Ø effective) rather than calculating the exact mathematical computed value of the MAE.

$$MAE = \frac{\sum_{i=0}^{n-1} |\Delta S|}{n} = \frac{\sum_{i=0}^{n-1} |S1-S2|}{n} \quad (1)$$

$$\emptyset_{effective} = \sum_{i=0}^{n-1} |\Delta S| \quad (2)$$

The calculated value of the effective discrepancy weight (Ø effective) from (2) will be compared with a predefined threshold value (Ø threshold) to determine whether the overall discrepancy values, over an adjustable time window, threaten or do not threaten the landing process. The APMU permits the adjustment of both the number of active memory elements and the Ø threshold. As illustrated in Fig. 6, we simulated the reading of the two sensors during a descending landing mission of a drone using MATLAB. The y-axis shows the scaled distance from the drone to the ground from the two sensors, and the discrepancy in their reading |Δ|. In this scenario, we could assume that the two readings are acceptable as they show approximately similar readings that refer to the descending landing approach of the drone as the two signal levels are reduced over time. However, if there is a malfunction occurs in one of the sensors, at any time interval, this could cause the absolute difference between the two signals |Δ| to be recorded as a noticeable discrepancy, as shown in the period from 3 to 4 for instance. This level of fluctuation of the |Δ| signal is an essential indicator to the pilot to decrease the dependability on the ALGAS4 system and should even switch to a semi-auto-landing mode instead. Thus, the ALGAS4 system provides an additional safety feature to the pilot. It is essential to mention that this proposed APMU, in this paper, has a more advanced key feature over the PMU in [5] by having the ability to adjust the effective window width (eww), which is shown in Fig. 6. The selection of the eww is also tuned based on the pilot opinion to whether having:

- A very sensitive ALGAS4 system toward any small |Δ| fluctuations by reducing the eww size. This could occur by deactivating a large group of memory storage elements. This could also cause a higher false alarm rate as well.
- A more mature sensitivity of the ALGAS4 system toward the |Δ| fluctuations by increasing the eww size. This could occur by activating a group of memory storage elements.
- Or, deciding to have a moderate response rate of the |Δ| fluctuations.

The values of the predefined Ø threshold are stored in an LUT and are configured based on the received signals from the FSAU. It is essential to realize that these values are left to the discretion of experts, who will vary and adapt them based on a variety of circumstances that are not the topic of this work. Also, the memory storage elements of the APMU consist of registers that hold the values while the ALGAS4 system is powered on. Based on the analysis of the proposed APMU module, its sensitivity correlates with both the operating frequency, Ø threshold, and the chosen event window size. As a result, achieving minimum sensitivity would require a minimum window of approximately 2.527ns to generate the specified output signal. However, under these conditions, the APMU's performance would mirror that of the PMU, resulting in the loss of its distinctive feature.

TABLE II
COMPARISON BETWEEN THE PROPOSED ALGAS4 SYSTEM AND OTHER PREDECESSOR ALGAS VERSIONS

|   | The Proposed ALGAS4 System in this paper | The Proposed ALGAS3 System in [5] | The Proposed ALGAS2 System in [4] | The Proposed ALGAS1 System in [11] |
|---|---|---|---|---|
| *FPGA device name* | INTEL 5CGXFC9D6F27C7 | | | |
| *FLS engines in the systems* | 4 systolic-basic cores + independent operation | | | 5 systolic-based cores |
| *Soft DSP units** | 2 FIR filters (15-TAPs) | | None | None |
| *Prognostic malfunction feature* | 16-bits configurable APMU | Fixed PMU | None | None |
| *Total dynamic and I/O thermal power dissipation* | 166.56 mW | 145.4 mW | 146.4 mW | 178.12 mW |
| *External memory usage* | None | None | None | None |
| *Junction temperature range* | 0 to 85 ºC | 0 to 85 ºC | 0 to 85 ºC | 0 to 85 ºC |
| *Selected cooling solution* | 23 mm heatsink with 200 LFpM airflow | | | |
| *Max. freq.* | 278.24 MHz | 276.63 MHz | 279.25 MHz | 266.03 MHz |
| *FLS crisp inputs resolution* | Variable (11-bits, 10-bits) | | | |
| *No. of logic elements* | 4,804 ALMs | 4,544 ALMs | 3,488 ALMs | 4,304 ALMs |
| *Maximum giga operations per second (GOPS)* | 74.847 GOPS | 70.82 GOPS | 21.22 GOPS | 25.273 GOPS |

\* Soft DSP units: DSP that has been implemented using the allocated ALMs and not the hard DSP block on the FPGA chip.

## IV. THE RESULTS

To maximize the computing performance of the proposed ALGAS4 processing core, the entire design was built from the ground up at the gate level and RTL using VHDL, with no IPs used. The results have also been verified by comparing the achieved results from the Questa Simulation tool and MATLAB, as has been explicitly discussed in previous contributions [4, 5, 11]. In this paper, we mainly focused on the performance of the proposed APMU and its comparison with the PMU from [5], as depicted in Table I. Even though the maximum frequency of the APMU is 8.18 percent lower than that of the PMU and the logic resources have increased by 3.32 times, the APMU provides the UAM drone with more flexibility and capabilities while consuming only 2.2 mW more dynamic and I/O power than the PMU. In addition, these findings were anticipated due to the APMU's coarse-grained architecture and the subsequent increased parallelism in computation. The computational performance of the proposed ALGAS4 system is compared with the predecessors' versions of it. According to Table II, the added functionalities to the ALGAS4 system will require around 4.8K ALMs of logic resources from the INTEL 5CGXFC9D6F27C7 FPGA device. This increase of approximately 5.41 percent in logic resources and 12.77 percent in total dynamic and I/O thermal power dissipation over the preceding ALGAS3 system is satisfactory. That is because the computational speed performance increased by around 1.05x of the ALGAS3 system, 3.52x of the ALGAS2 system, and 2.96x of the ALGAS1 system.

This computational improvement of the proposed ALGAS4 system is due to the additional systolic subblocks of the proposed APMU, which are firing their outputs simultaneously after each clock cycle. Additionally, the emphasis on dynamic power consumption in this paper is due to the need for meaningful data that will be used in future comparisons with similar ASIC-based systems. The static power consumption of FPGA-based designs represents the entire static power consumption per chip and not per the implemented system. In addition, Table II presents further countermeasures that have been considered for defending the ALGAS4 system against cyberattacks and making it more durable to withstand and recover from error conditions during any fault injection attacks by avoiding the requirement to interchange data with any sort of external memory while processing the input sensory data from the two sensors. This feature is achieved by storing all the required coefficients, parameters, and weights for processing all the different data stages of the ALGAS4 architecture using internal soft registers within the silicon fabric of the FPGA. This also includes the FIFO-like memory storage buffers, which are used to implement the APMU. Subsequently, the usage of any external memory or embedded BRAM cells is zero. Furthermore, this independence from external memory cells enables the ALGAS4 system to achieve a memory access time of zero. Such design consideration is an essential key element to achieve the high computational speed for the ALGAS system in general.

TABLE III
COMPARISON BETWEEN THE PROPOSED ALGAS4 SYSTEM AND OTHER SIMILAR SAFE AND PRECISE LANDING SYSTEM SYSTEMS

|  | The Proposed ALGAS4 System in this paper | The Safe and Precise Landing System in [22] | The Safe and Precise Landing System in [23] | The Safe and Precise Landing System in [24] |
|---|---|---|---|---|
| *Main algorithm(s) to support safe landing* | Fuzzy logic system, and other DSP units | Passive system (Send the aggregated sensory data to the main controller) | Kalman filter, linear least squares (LLSs) method, and other DSP units | Computer vision using the depth camera |
| *Processing hardware platform* | - 4x INTEL 5CGXFC9D6F27C7 FPGAs | - Xilinx Multi-Processing System on a Chip (MPSoC):<br>- 4x ARM A53 processors<br>- 2x ARM R5 real-time processors<br>- FPGA | - NVIDIA Jetson TX2 processor | - IntelM3-8100Y 3.4GHz dual-core processor |
| *Sensors used for the safe landing* | - Four 24 GHz radar sensor<br>- Four 840nm Lidar sensor | - Inertial Measurement Unit (IMU)<br>- Navigation Doppler LIDAR (NDL)<br>- A camera for Terrain Relative Navigation (TRN)<br>- A Hazard Detection LIDAR (HDL) | - Inertial measurement unit (IMU), UWB tags | - Intel RealSense D435i camera |
| *Redundancy feature* | 4 systolic-basic cores + independent operation | No | No | No |
| *Porting and migration flexibility* | Low | Intermediate | High | High |
| *Cross-platform development and reimplementation* | Only within FPGA platforms | Flexible within CPU/FPGA-based platforms | Flexible within CPU-based platforms | Flexible within CPU-based platforms |
| *Avoid depending on intellectual property (IP) circuits* | Yes | No | No | No |

Also, the ultra-low dynamic power consumption of the ALGAS4 system aligned with a similar family of ALGAS systems as concluded in Table II. This feature is a key aspect since the main targeted applications of the proposed ALGAS family modules are in mobile devices and systems. As depicted in Fig. 3, the ALGAS4 system consists of many processing elements. Each of these elements has its own clock domain and control signals. Thus, another key element in enhancing synchronization among these modules within ALGAS4 system involves selecting suitable synchronization methods. These methods guarantee that the modules function cohesively and uniformly alongside the entire system. In fact, we used a combination of different synchronization mechanisms including clock signals, handshake signals, interrupts, and semaphores. They were assigned based on the group of technical rules to sustain the operational efficiency of the ALGAS4 system.

Detailing the adjusted approaches to enhance FPGA area utilization in this design is crucial. Among various strategies accessible for employment, the proposed ALGAS4 system emphasis was directed solely towards two: the considerations regarding arithmetic operations and the selection of data types. Each module within the proposed systolic ALGAS4 system was customized to precisely suit its designated function. For instance, we steered clear of employing arithmetic operations reliant on floating-point data types due to their considerable power consumption. Moreover, these data types necessitate a greater number of FPGA cells for implementation. Also, we constructed an intricate bus network to interconnect the diverse modules within the ALGAS4 system. Each bus's width was fine-tuned based on the maximum value of data handled at each stage of the design for optimal performance. These employed utilization techniques formed the cornerstone for boosting the computational performance of the ALGAS4 system. Yet, they resulted in a maximum degradation of the ALGAS4 system's output accuracy by approximately ±5% compared to the values tested using MATLAB.

Despite implementing various logic utilization techniques, it's essential to note that the updated version of the ALGAS system, the ALGAS4, occupies a larger area compared to its predecessors, as shown in Table II. This expansion primarily stems from the necessary logic modules required for implementing the new feature, the 16-bit configurable APMU. Also, the proposed ALGAS4 system significantly occupies fewer logic cells compared to the state-of-the-art systems listed in Table III. This is attributed to its sole dependence on FPGA platforms, without involving any general-purpose CPU.

To emphasize the main advantages of the ALGAS4 system over the state-of-the-art systems, we summarize its key aspects, as depicted in Table III. Despite the impressive technologies and implemented ideas that have been used in the References [22, 24], they mainly depend on cameras to enhance safe drone landing mechanisms. This is effective only for indoor environments, but it is not suitable for most of the outdoor landing situations. Reference [23] tries to avoid the usage of cameras, but it needs a visual human to make sure that the drone is close to the anchor platform before the auto-landing mechanism starts to perform its operation.

It is also essential to mention that the proposed ALGAS4 system is not based on a heterogeneous architecture and has been designed from scratch without depending on any Intellectual Property (IP) circuits. This is an extremely important feature in terms of enhancing the security of the hardware, boosting the computational speed, and optimizing the overall power consumption of the ALGAS system. This is mainly because these commercial IPs are designed to fit into a wide range of applications regardless of their constraints.

In addition to the APMU safety factor, the ALGAS4 system is enhancing the safety factor by depending on four spatially separated FPGAs and it has the ability to keep functioning tolerably under the failure of any pair of these FPGAs under the condition that the drone's pilot permits this case.

Conversely, to ensure fairness, we conducted a comparison with other state-of-the-art systems, considering various key system features such as:
•	Porting: refers to the process of transferring a design from one platform to another, ensuring its functionality remains unchanged.
•	Migration: involves relocating a design or system from one platform to another, frequently necessitating modifications to hardware-specific features or configurations.
•	Cross-platform development: involves crafting a design capable of seamless operation across diverse hardware platforms or architectures.
•	Reimplementation: involves reconstructing or remaking a design to align with the structure and limitations of an alternate hardware platform.

As depicted in Table III, the presented ALGAS4 system demonstrates limited performance in meeting these criteria, as it's solely applicable within FPGA-based platforms and cannot be transferred elsewhere.

## V. THE CONCLUSION

This paper introduced the APMU feature to the ALGAS4 system, which is vital for boosting the SESAR and NextGen safety measures during the autonomous landing mechanism of the futuristic taxi drone. Due to the APMU, the computing performance of the ALGAS4 system has reached 74.85

GOPS. The current adjustments to the ALGAS4 architecture will serve as the foundation for future iterations of this architecture to achieve more tangible processing performance that targets FPGA and ASIC platforms.

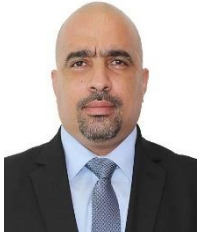

**Hossam O. Ahmed** became a Member (M) of IEEE in 2014, and an IEEE Senior Member (SM) in 2021. He received the MSc and PhD in Electronics and Electrical Communications Engineering from Ain Shams University, Egypt, in 2015 and 2019, respectively. He is currently working as an assistant professor in the computer engineering department at the American University of the Middle East (AUM), Kuwait. He is interested in wireless sensor network applications, artificial intelligence hardware accelerators, intelligent control systems, fuzzy systems applications, and control systems design using VHDL/FPGA. Also, he is an active member of the SASD Voting Members (Standards Activity Subdivisions Committee (SASD)). He is an active reviewer for IEEE Transactions on Circuits and Systems I; IEEE Transactions on Circuits and Systems II; IEEE Transactions on Consumer Electronics, IEEE Transactions on Computer-Aided Design of Integrated Circuits and Systems, IEEE Access, IEEE Computer Architecture Letters, and Elsevier Microprocessors and Microsystems, the IEEE International Symposium on Circuits and Systems (ISCAS), and the IEEE/ASME International Conference on Advanced Intelligent Mechatronics (AIM).

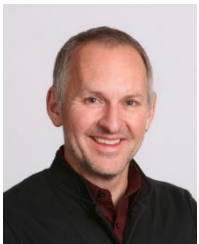

**David Wyatt** is the CTO/President of both PixelDisplay and CardWare Inc, companies that he founded to realize technology solutions to valuable problems. David's career has encompassed a wide range of disruptive innovations and is the named inventor on more than 140 issued US patents. David originates from rural Australia, the country where he studied Computer Science at the University of Queensland, and Electrical Engineering at South Brisbane College, before following his passion for both hardware and software engineering to Taiwan. And then emigrating to the US, through a Silicon Valley acquisition. With leading roles as Chief Engineer & Platform Architect at Intel for 8yrs, and Distinguished Engineer at NVIDIA for 9yrs. David now resides in Austin Texas, aka Silicon Hills, where he's hard at work on the next generation OpenGPU™ architecture, amongst other ventures.